# A Transformer Based Generative Chemical Language AI Model for Structural Elucidation of Organic Compounds

*Xiaofeng Tan*

X Scientific, 1 Bramble Way, Acton, MA 01720. Email: x.tan@jhu.edu

ABSTRACT

For over half a century, computer-aided structural elucidation systems (CASE) for organic compounds have relied on complex expert systems with explicitly programmed algorithms. These systems are often computationally inefficient for complex compounds due to the vast chemical structural space that must be explored and filtered. In this study, we present a proof-of-concept transformer based generative chemical language artificial intelligence (AI) model, an innovative end-to-end architecture designed to replace the logic and workflow of the classic CASE framework for ultra-fast and accurate spectroscopic-based structural elucidation. Our model employs an encoder-decoder architecture and self-attention mechanisms, similar to those in large language models, to directly generate the most probable chemical structures that match the input spectroscopic data. Trained on ∼ 102k IR, UV, and ¹H NMR spectra, it performs structural elucidation of molecules with up to 29 atoms in just a few seconds on a modern CPU, achieving a top-15 accuracy of 83%. This approach demonstrates the potential of transformer based generative AI to accelerate traditional scientific problem-solving processes. The model's ability to iterate quickly based on new data highlights its potential for rapid advancements in structural elucidation.

# 1. Introduction

For over half a century, various spectroscopic techniques have been developed for identifying organic compounds, each providing unique insights into molecular structures [1]. The structural elucidation of organic compounds is crucial for understanding their physical properties, chemical behavior, reactivity, and applications in various fields such as pharmaceuticals, materials science, chemistry, and biological sciences. Spectroscopic techniques offer detailed structural information that is essential for comprehensive molecular characterization. Infrared (IR) spectroscopy is used to determine the functional groups within a molecule by measuring the vibration frequencies of chemical bonds, which are primarily influenced by the functional groups they belong to [2]. Ultraviolet-visible (UV-Vis) spectroscopy is useful for understanding electronic transitions and conjugation within organic molecules [3]. 1D and 2D Nuclear Magnetic Resonance (NMR) spectroscopy, both proton ($^1$H NMR) and carbon ($^{13}$C NMR), provide detailed information about the molecular framework, including the environment around specific nuclei [4,5]. Mass spectrometry (Mass Spec) offers precise molecular weight determination and fragmentation patterns that help elucidate the structure of organic compounds [6]. Despite their power, these techniques pose challenges such as the complexity of spectra, overlapping signals, and the need for expert interpretation.

The development of Computer-Aided Structure Elucidation (CASE) systems began in the latter half of the 20th century, driven by the motivation of using computers to handle large amounts of spectroscopic data automatically and accurately [7–10]. Traditional structure elucidation process involves integrating data from various aforementioned spectroscopic techniques to infer molecular structures. CASE systems enhance this process by employing extensive reference libraries and logical-combinatorial algorithms, leveraging highly sophisticated expert systems to generate structural hypotheses. Early research, despite not succeeding with complex molecules due to limited 1D NMR data, laid the groundwork for CASE methodologies and core algorithms, establishing a strategic framework that continues to evolve. Pioneering works and subsequent reviews have documented these advancements, underscoring both the achievements and challenges in achieving a fully automated structure elucidation workflow [11–25].

The common workflow of a CASE system begins with the acquisition of molecular spectra that serve as the initial data for the analysis. Positive structural constraints, mainly molecular fragments, are determined from these spectra and input into the system. Some systems, such as ACD/Structure Elucidator [25], generate a molecular connectivity diagram (MCD) and assemble structural fragments into candidate structures, which are then ranked based on their agreement with experimental data [11,22]. Using this information and the molecular formula that is usually obtained from high resolution Mass Spec data, the system generates possible isomers, ensuring that only those meeting all imposed constraints and spectral features are considered. The system uses generation algorithms to exhaustively produce all structures that satisfy these conditions. Spectral filters, including typical molecular fragments and their corresponding spectral features, are applied to verify the compliance of the generated structures with the experimental spectra. Any structures containing fragments not confirmed by the spectra are excluded from the final output. The process involves checking the proximity of calculated and experimental spectra, selecting the structure whose predicted spectra most closely match the experimental data. This method, although computationally intensive, particularly when generating and filtering numerous potential structures, is critical to ensure a high level of accuracy in identifying the most probable molecular structure.

The fundamental framework of CASE systems has remained largely unchanged over the past 50 years although its individual algorithms have evolved significantly. For example, with the integration of new NMR experiments and computational methods, such as density functional theory (DFT), the accuracy and robustness of CASE systems has been enhanced significantly [14,15]. New generations of CASE systems, such as DP4-AI [26], also employ automated algorithms to significantly improve efficiency of structural elucidation. Although increased computer processing power and automated algorithms have accelerated structure generation and filtering calculations, these two steps still form speed bottlenecks in the CASE workflow, particularly for complex molecules. For complex molecules, the number of structures needed to be generated and filtered could easily reach tens to hundreds of thousands or even more, causing the CASE framework to require minutes or even hours to solve the structural elucidation problem [11,12].

Recent advancements in deep learning (DL) and generative AI have revolutionized various scientific domains and shifted scientific expert systems from traditional explicit programming paradigm towards data-driven and end-to-end DL approaches. In image recognition, DL models, particularly convolutional neural networks (CNNs), have achieved superhuman performance in tasks like object detection and facial recognition [27,28]. Natural language processing (NLP) has seen similar breakthroughs with transformer-based [29] large language models (LLM) like BERT [30] and GPT [31], enabling machines to understand and generate human language with remarkable accuracy. AlphaFold [32], developed by DeepMind, represents a significant leap in predicting protein structures, solving a long-standing challenge in biology and offering new insights into biomolecular interactions. Compared to these advancements, the adaptation of AI in CASE systems has been relatively slower, with initial efforts focused on improving individual algorithms within the existing expert system. For example, AI has been employed for quick analysis of NMR spectra [7,33–36], and significant development has occurred in AI models for fast DFT calculations, driven primarily by their applications in electronic structure analysis and calculations.

Over the past three years, AI has been directly applied to molecular structural elucidation. Skinnider *et al.* employed a recurrent neural network (RNN) based generative model for molecular structure elucidation, though their approach has been primarily limited to new psychoactive substances due to the necessity of having a structural prior [37]. In another development, Tian *et al.* introduced a CNN-based binary classification model, MatCS, which directly predicts the relationship between NMR spectral images and the molecular structure of the target compound [38]. The most significant advancements, however, come from recent work by Alberts *et al.*, who employed transformer-based sequence-to-sequence models to predict molecular structures using NMR and IR spectroscopic data [39,40].

Inspired by the great success of LLMs in text-to-text generation, and similar sequence-to-sequence generative AI models such as image captioning [41], prompt-based image generation [42,43], and video generation [44], we developed a generative chemical language model for the structural elucidation of organic compounds (CLAMS) using their spectroscopic data. This model aims to provide the functionality of the conventional expert system in CASE but replaces explicitly programmed algorithms with a transformer-based encoder-decoder architecture for structural generation and elucidation. Although CLAMS and the model developed by Alberts et al.[39,40] both employ transformer-based architectures, our work was conceived and developed entirely independently. The CLAMS model addresses specific challenges in the integration of spectroscopic data with chemical structure prediction, incorporating distinct design choices and probably more importantly datasets. In this paper, we present the architecture of CLAMS and preliminary testing of this novel system for fully automated spectroscopic based structural elucidation of organic compounds. The paper is organized as follows: Section 2 presents the architecture of the CLAMS model, along with the datasets and training procedures. Section 3 provides benchmarks of CLAMS for functional group classification and structure elucidation. Finally, Section 4 offers conclusions and discussions.

## 2. Methods and Data

### 2.1. The CLAMS Model

The CASE system can be viewed as a sequence-to-sequence translation system, where spectroscopic data serve as the input sequence and molecular structures as the output sequence. For a DL model to function effectively as a CASE system, molecular data must be encoded into representations that computers can process efficiently. Various molecular encodings or featurizers exist, such as atomic coordinates, molecular graphs, molecular descriptors, fingerprints, distance matrices, Coulomb matrices, and SMILES, each with unique advantages and disadvantages [45,46]. To qualify as an effective molecular featurizer for structural elucidation in a DL model, the featurizer must uniquely determine molecular structures and be efficiently usable in a sequence-to-sequence translation model. In CLAMS, we employ SMILES [47–49] as the molecular featurizer. SMILES notation encodes molecular structures as text strings by representing atoms and bonds in a linear format, enabling easy storage and manipulation. This textual format allows the application of powerful transformer-based DL architectures to efficiently analyze chemical information.

The architecture of our CLAMS model is illustrated in Fig. 1. The model employs an encoder-decoder architecture for translating spectroscopic data into SMILES strings that encode molecular structures. The CLAMS model utilizes a Vision Transformer (ViT) [50] as its encoder, comprising 9 hidden layers, each with 9 attention heads. The input to the ViT encoder is constructed by sequentially concatenating 1D spectroscopic data from IR, UV-Vis, and 1D ¹H NMR spectroscopy into a single 1D array of 4356 elements. This array is reshaped into a 66 x 66 image format. Mass Spec data is not available in the dataset we used to train and test the CLAMS model. The reshaped image is divided into 11 x 11 patches, each consisting of 6 x 6 pixels. Each of the 121 patches is processed by a 2D convolutional layer with a kernel size of 6 x 6 and 288 feature maps, transforming them into 1D feature vectors, or patch embeddings, of dimension 288. The kernel size is set to match the patch size so that a single 2D convolutional layer can capture global features within each patch. These 121 patch embeddings are then combined with positional embeddings, which capture the positional relationships among the patches, to form the input for the ViT encoder. The input data is processed through 9 residual self-attention layers, extracting complex features among different spectroscopic data by weighing their importance relative to each other through the self-attention mechanism [29].

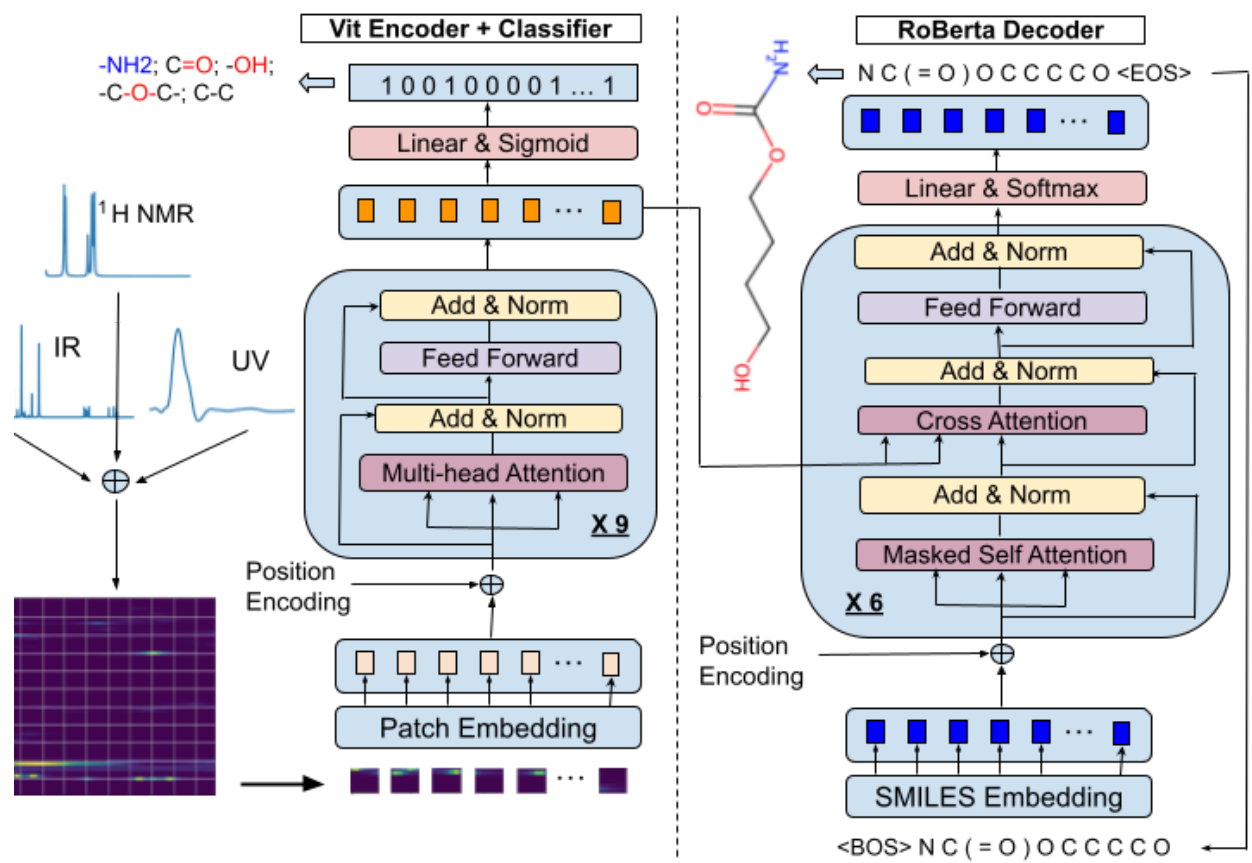

**Figure 1.** Schematic of the CLAMS model architecture. The left side of the dividing line features the ViT encoder and the 18-functional-group MLP classifier, with an example label illustrating detected functional groups. The right side showcases the ChemBERTa decoder, including an example of a generated SMILES string, demonstrating the iterative sequence generation process with tokens displayed above and below the decoder. The SMILES string uses <BOS> (begin-of-sequence) and <EOS> (end-of-sequence) tokens. Previously generated tokens (top) are used by the encoder to help generate the current token (bottom). The corresponding molecular structure is shown to the left of the decoder. Solid filled rectangles represent embeddings/features at different stages of processing within the model. The numbers of hidden layers in the encoder and decoder are labeled with underscore text in the figure.

The data features extracted from the ViT encoder are routed to two distinct networks. The first network is a fully connected multi-layer perceptron model (MLP) with a sigmoid output layer comprising 18 outputs, functioning as a multi-label classifier. This classifier determines the presence of 18 functional groups (i.e., alkane, alkene, alkyne, arene, haloalkane, alcohol, aldehyde, ketone, ester, ether, amine, amide, nitrile, imide, thial, phenol, enamine, carbamate) within the molecule based on its input spectroscopic data and is used only to pretrain the ViT encoder. The second network is a pre-trained RoBERTa model [51] called ChemBERTa by Liu *et al.* [52,53], which serves as the decoder for the CLAMS model. The ChemBERTa was pre-trained to learn molecular fingerprints encoded in SMILES strings through a semi-supervised training on a dataset containing approximately 10 million PubChem [54] compounds. The ChemBERTa decoder features 6 layers of RoBERTa attention layers. The sequence of features extracted by the ViT encoder is fed into the RoBERTa decoder, where it undergoes layers of self-attention and feed-forward operations to capture contextual information of chemical structures associated with the data. Structural elucidation is achieved in the process of model generation of SMILES strings in which each token in the string is predicted sequentially, with the model utilizing previously generated tokens as part of its input context. This iterative process begins with the begin-of-sequence (<BOS>) and continues until the end-of-sequence token (<EOS>) is generated, resulting in a coherent sequence of output tokens that form the decoded SMILES output. This output SMILES string specifies the most probable chemical structure that matches the input spectroscopic data.

**2.2. Data Preparation**

We prepared the datasets for training and testing our CLAMS model using the QM9S dataset constructed by Zou et al.[55]. Derived from the QM9 dataset[56,57], QM9S includes 129,817 organic molecules composed of C, F, N, O, and H, with the largest molecule in the dataset containing 9 heavy atoms and 29 total atoms. Table 1 shows the size distribution of the molecules in the dataset. Compared to QM9, QM9S incorporates molecular properties calculated using DFT at the B3LYP/def-TZVP level of theory, including tensors of various ranks such as energy, partial charges, electric dipole, Hessian matrix, quadrupole moment, polarizability, octupole moment, and first hyperpolarizability. QM9S was created to demonstrate the capabilities of DetaNet, an E(3)-equivariant DL network developed by the authors, in calculating these properties at quantum chemical accuracy with a speed 2 – 4 orders of magnitude faster than DFT.

Using the molecular properties in QM9S, we calculated the IR, Raman, UV–Vis, and 1D $^1$H NMR spectra for all the molecules in QM9S. Molecules with invalid (i.e., NANs) spectral values were removed, resulting in a spectroscopic dataset of 127,465 molecules. This dataset was randomly divided into three subsets using a random splitter: 80% for training, 10% for validation, and 10% for testing, maintaining the same size distribution as QM9S. Each IR spectrum contains 3501 data points, zero-padded to 3600, expressed as absorbance in the 100 – 4000 cm$^{-1}$ range. It was calculated using fundamental harmonic frequencies, broadened with a Lorentzian profile (FWHM of 5 cm$^{-1}$), and max-normalized to [0, 1]. We found that the CLAMS model performs better with IR absorbance spectra than with transmittance spectra, likely due to information loss when strong absorption causes near-zero transmittance values, as we had hypothesized. The UV-Vis spectrum spans 100 – 400 nm with 240 data points, intensity max-normalized to [0, 1]. The 1D $^1$H NMR spectrum covers the 0 – 12 ppm chemical shift range with 516 data points, broadened with a Lorentzian profile (FWHM of 0.05 ppm) and max-normalized to [0, 1]. Our generated dataset does not contain Mass Spec data as it cannot be easily calculated from the available molecular properties.

**Table 1. Size distribution of the QM9S dataset categorized by the number of heavy atoms in each molecule.**

| No. Heavy Atoms | Count | Smallest Molecular Size | Largest Molecular Size | Median Molecular Size |
|---|---|---|---|---|
| 1 | 3 | 3 | 5 | 4 |
| 2 | 5 | 3 | 8 | 4 |
| 3 | 9 | 6 | 11 | 7 |
| 4 | 28 | 5 | 14 | 9.5 |
| 5 | 123 | 6 | 17 | 11 |
| 6 | 588 | 7 | 20 | 13 |
| 7 | 3052 | 8 | 23 | 15 |
| 8 | 17442 | 9 | 26 | 16 |
| 9 | 106215 | 10 | 29 | 18 |

## 2.3. Model Training

We trained the CLAMS model in two steps. First, we pre-trained the ViT encoder alongside the MLP multi-label classifier using a supervised learning approach. Spectroscopic data and functional-group labels were input into the ViT + MLP classifier sub-network to optimize the system's weights, enabling it to map spectroscopic data to molecular functional groups. This process allowed the ViT encoder to learn an effective representation of the input data. In the second step, we connected the ViT encoder to the pre-trained ChemBERTa decoder and fed it with spectroscopic data and SMILES strings for supervised training of the entire CLAMS model. In this step, the MLP sub-network was not used. The aim was to optimize the pre-trained weights of both the ViT encoder and the ChemBERTa decoder for accurate structural generation and elucidation.

During the pre-training phase, we performed a light hyper-parameter optimization on the ViT encoder in the CLAMS model. This involved a grid search to determine the optimal number of attention heads per hidden layer, the number of hidden layers, and the hidden size. Two constraints were applied to reduce the search space: the number of attention heads per hidden layer was set equal to the number of hidden layers, and the hidden size was fixed at 32 times the number of attention heads. The number of hidden layers varied between 6 and 12. We selected the best parameters based on the classification accuracy of the ViT + MLP sub-network for the 18 molecular functional groups, resulting in an optimal configuration of 9 hidden layers, 9 attention heads, and a hidden size of 288 of the ViT encoder.

To ensure no information leakage during model testing, we used the same data split for pre-training, training, and testing. The training subset was utilized for pre-training the ViT encoder and training the full CLAMS model. The validation subset helped select the best models during both the pre-training and training, while the test subset evaluated the ViT + MLP classifier and the final CLAMS model. We employed early stopping and dropout techniques to prevent overfitting. The entire CLAMS model, excluding the MLP classifier, comprises 104,038,624 trainable parameters.

3. **Results**

In this section, we evaluate the performance of our CLAMS model in functional group classification and structural elucidation. As described in Section 2, we generated a dataset of over 127,465 organic compounds. This dataset was divided into training, validation, and test subsets to facilitate the model's training and evaluation.

### 3. 1. Functional-Group Classification

In a comparative study, we evaluated the performance of the ViT + MLP classifier sub-network in classifying the 18 functional groups using two sets of spectroscopic data. In the first experiment, we trained and tested the ViT + MLP classifier using the IR, UV-Vis, and 1D $^1$H NMR spectra in our training subset, concatenated into 1D arrays of 4356 points and reshaped into 66 x 66 images. In the second experiment, only the IR data in our training subset, reshaped into 60 x 60 images, were used. This study aimed to gain information on how the inclusion of additional spectroscopic data types affects the performance of the CLAMS model, providing insights into the potential improvements that could be achieved with the addition of Mass Spec data, which is routinely used in structural elucidation of organic molecules but was unfortunately not available in our dataset.

Fig. 2 illustrates the performance metrics of the ViT + MLP classifier in classifying the 18 molecular functional groups in the test subset. Fig. 2(B), 2(C), and 2(D) respectively display the classification precision, recall, and F1 score for each functional group under the two different experimental conditions. The ViT + MLP classifier demonstrates strong performance in functional group identification, especially when trained exclusively with IR data, achieving a minimum F1 score of 0.92 across all the 18 functional groups. The results indicate that using only IR data improves classification accuracy, as the inclusion of UV-Vis and 1D $^1$H NMR data introduces additional data that complicates model optimization but without providing additional information for molecular functional group classification beyond what the IR data already provides.

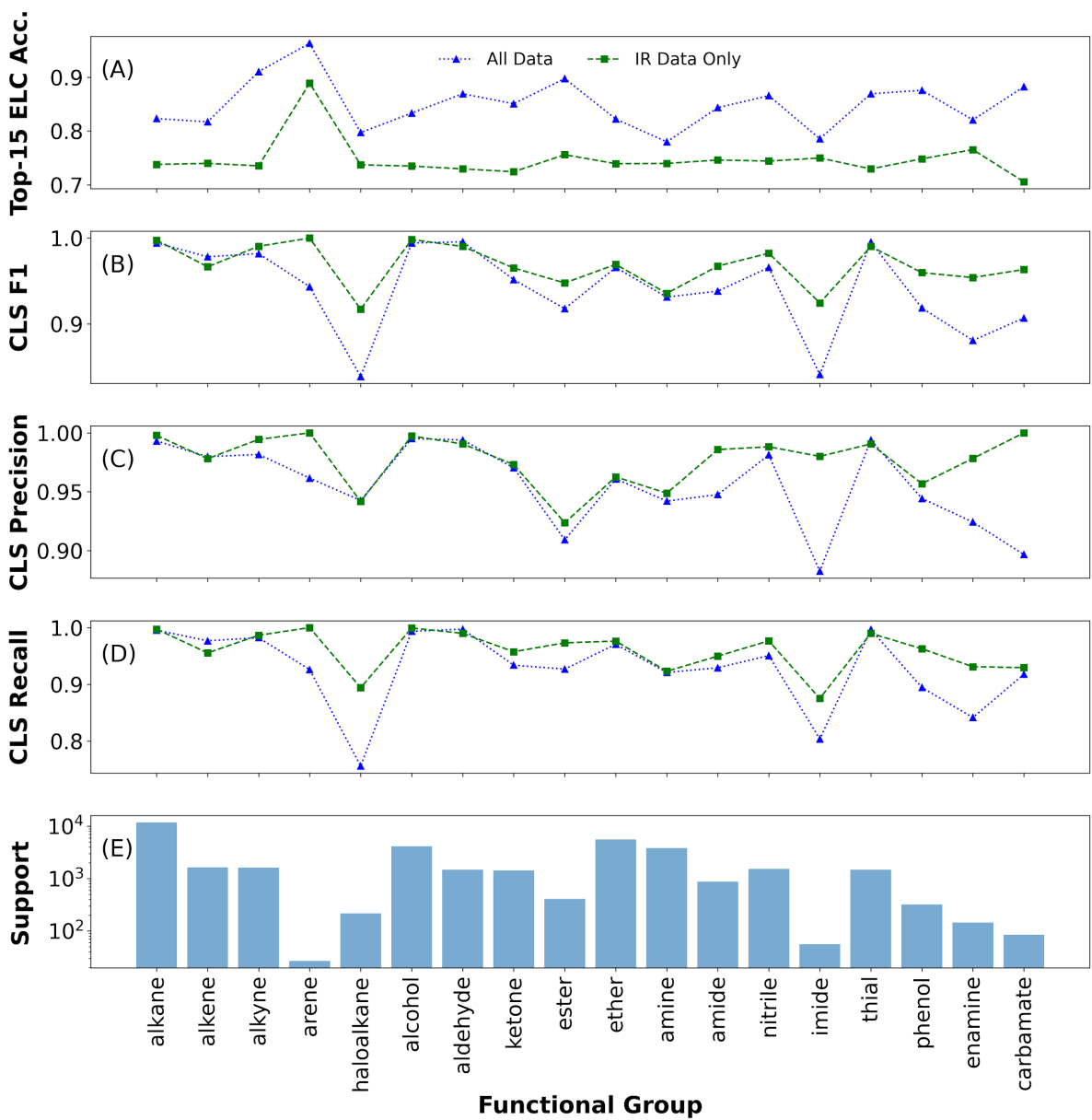

**Figure 2.** Model performance metrics of the ViT + MLP model and the entire CLAMS model, broken down for each functional group. (A) Top-15 structural elucidation accuracy. (B) Functional group classification F1 score. (C) Functional group classification precision score. (D)

Functional group classification recall score. (E) Support number (i.e., number of occurrences) of the 18 functional groups. For (A) through (D), metrics for the model trained on all spectroscopic data (IR + UV-Vis + 1D $^1$H NMR) are shown as triangles connected with dotted lines, while metrics for the model trained with IR data only are shown as squares connected with broken lines.

### 3.2. Chemical Structural Elucidation

As detailed in the Materials and Methods section, the training of the CLAMS model began with pre-training the ViT + MLP sub-network for functional group classification within the dataset. After that the ViT encoder was integrated with the ChemBERTa model, which had been pretrained on 10 million PubChem organic compounds. The resulting CLAMS model was subsequently trained using the same training subset employed for the ViT + MLP pre-training. Once fully trained, the structural elucidation process with CLAMS functions similarly to prompt-based text generation in an LLM model. Here, the concatenated spectroscopic data serves as the prompt, and the CLAMS model generates a SMILES string representing the most probable chemical structure associated with the input data. Once trained, the structural elucidation process or the SMILES generation process in the CLAMS model is highly efficient, taking only a fraction of a second to produce a single SMILES string, even on a CPU.

To increase the likelihood of accurately identifying the correct chemical structure, it is beneficial for the CLAMS model to generate the top-$k$ chemical structures (k ≥ 1) simultaneously using a beam search algorithm [58]. Beam search explores multiple potential sequences at each step, maintaining the top $k$ sequences based on their probabilities. CLAMS employs aggressive beam search, which ensures thorough exploration by consistently selecting the most probable beams, enhancing model performance. In contrast, sampling-based beam search introduces randomness by sampling from the probability distribution, useful for text generation in NLP but not optimal for chemical structure generation where accuracy is paramount.

We evaluated the structural elucidation accuracy of CLAMS using the test subset with an aggressive 15-beam search to generate top-1, top-5, top-10, and top-15 structures. For each top-$k$ evaluation, success was determined if any of the $k$ generated SMILES, after canonicalization, matched the canonicalized ground-truth SMILES. Similarly to how we tested the ViT + MLP sub-network, we separately evaluated the CLAMS model using all IR, UV-Vis, and 1D $^1$H NMR data compared to using only IR data. Fig. 2(A) demonstrates that the full CLAMS model, trained with IR, UV-Vis, and 1D $^1$H NMR data, outperforms the model trained with IR data alone. This highlights the importance of UV-Vis and 1D $^1$H NMR data in structural elucidation of organic compounds as they provide additional critical information on functional group arrangement, despite not adding to the functional group identification beyond what IR data provides. Fig. 3 extends this by illustrating the overall (i.e., not categorized by functional groups) structural elucidation accuracy as a function of the number of generated structures, $k$, for both CLAMS model variants. To offer more comprehensive insight into the quality of the generated structures, we also present Murcko scaffold accuracy[59] alongside structural elucidation accuracy in Fig. 3.

Fig. 3 shows that increasing the number of generated sequences improves the likelihood of correctly identifying the target structure. However, scaffold accuracy decreases as the number of generations increases, since the order of the generated structures in the CLAMS models reflects the relative probability of matching the true structure. Consequently, the improvement in structural elucidation accuracy slows significantly when the number of generations exceeds 15. It is also important to note that while the CLAMS models generally generate valid SMILES, the ratio of valid SMILES is not always perfect, though it remains close to 1, especially for smaller *k* values. Fig. 4 presents the valid generation ratio as a function of the number of generated structures, *k*, illustrating that this ratio declines as *k* increases, with a sharper drop when *k* exceeds 10. Similar to the trend seen in functional group classification accuracy, the CLAMS model trained solely on IR data produces a higher ratio of valid SMILES compared to the model trained with IR, UV-Vis, and 1D $^1$H NMR data.

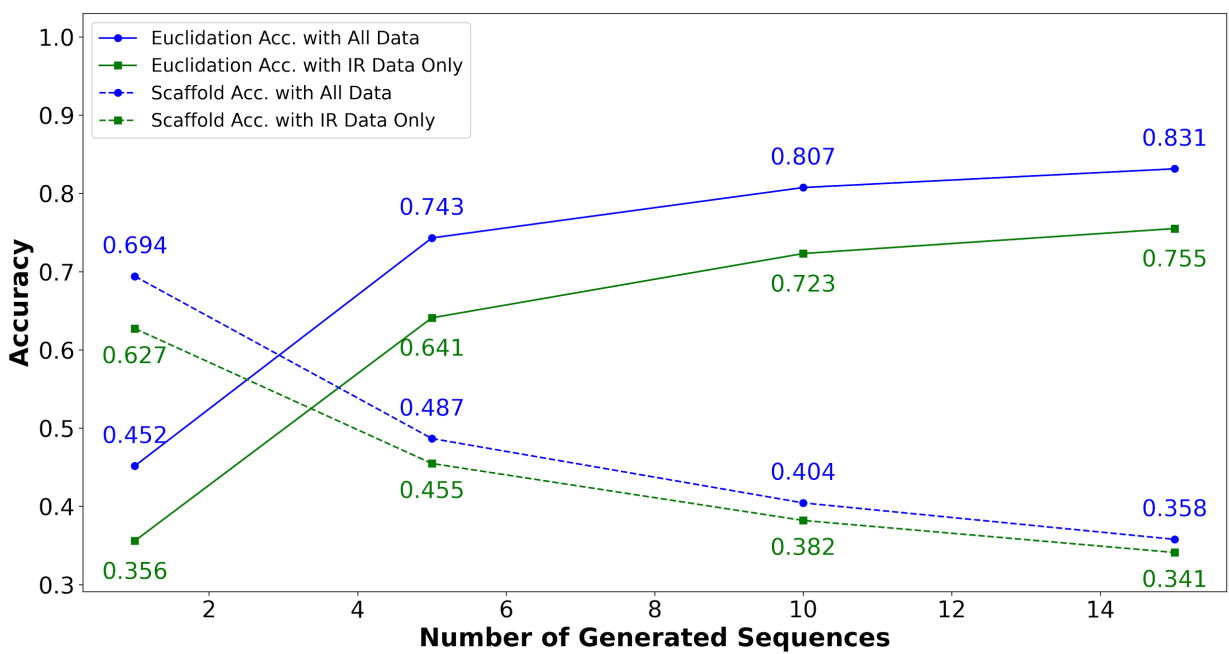

**Figure 3.** Structural elucidation accuracy and scaffold accuracy as functions of the number of generations in the CLAMS model. The blue curves represent the model trained on all data (IR + UV-Vis + 1D $^1$H NMR), while the green curves represent the model trained solely on IR data. Both metrics are shown for each top-k generation.

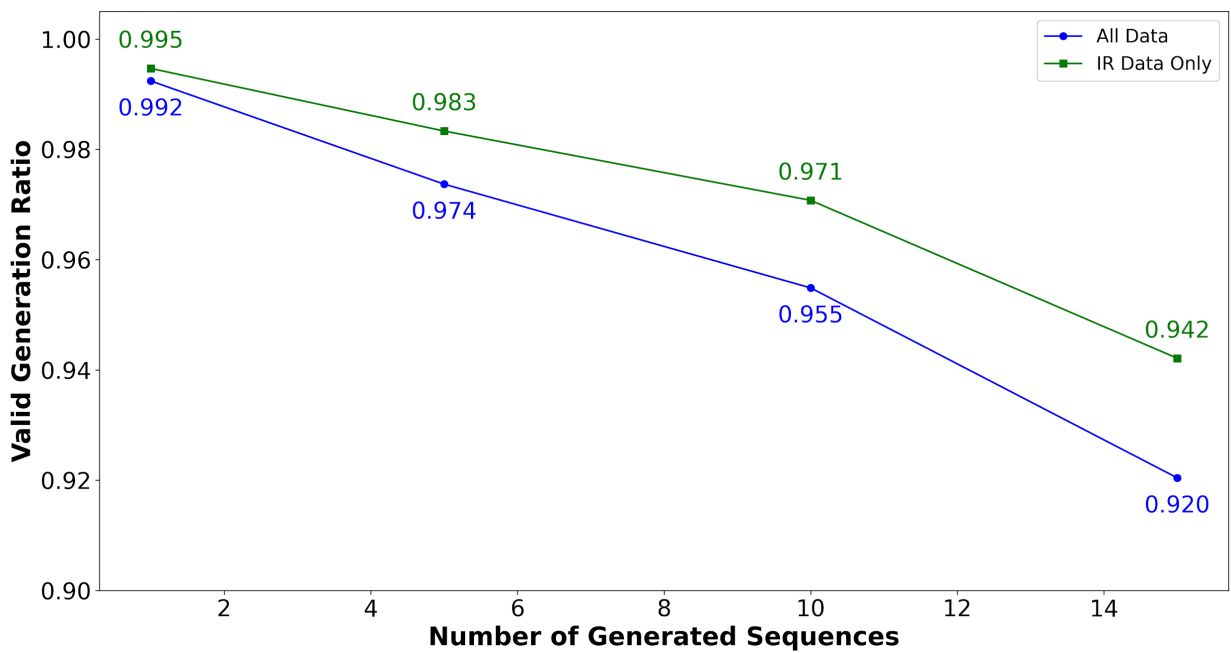

**Figure 4.** Valid generation ratio as a function of the number of generations in the CLAMS model. The blue curve represents the model trained on all data (IR + UV-Vis + 1D $^1$H NMR), and the green curve represents the model trained solely on IR data.

Fig. 5 showcases several top-5 structural elucidation examples using the CLAMS model. In Fig. 5(A), CLAMS correctly identifies the structure of the compound, but with different SMILES notations than those in the test subset. This occurs because a molecule can have multiple valid SMILES representations or synonyms. While training on canonicalized SMILES can resolve this synonym issue, we found that the CLAMS model trained on non-canonicalized SMILES outperformed the canonicalized approach, as shown in Table 2. As a result, all performance metrics reported here, except some of those in Table 2, are based on the model trained without SMILES canonicalization. In Fig. 5(B), CLAMS generates the correct structure twice with matching SMILES as in the test subset. In Fig. 5(C), CLAMS generates the correct structure twice but with different SMILES as in the test subset. In Fig. 5(D), CLAMS fails to identify the correct structure in the top-5 elucidation process.

(A)  Truth: CCCC1=NNC(N)=N1    Pred#1: CCCc1n[nH]c(N)n1    Pred#2: CCCCc1n[nH]c(N)n1

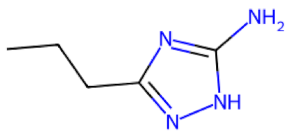 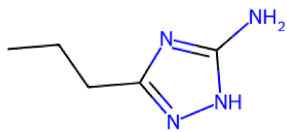 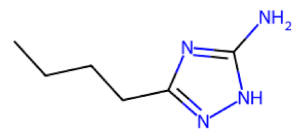

Pred#3: CCCc1nc(N)n[nH]1    Pred#4: CCCNc1n[nH]c(N)n1    Pred#5: CCCCc1nc(N)n[nH]1

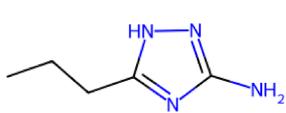 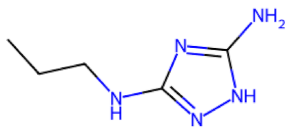 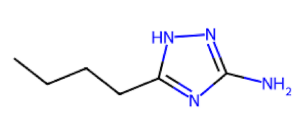

(B)  Truth: CNC(=O)N(C)C(C)C    Pred#1: CCN(C)C(=O)NC    Pred#2: CCC(=O)N(C)C

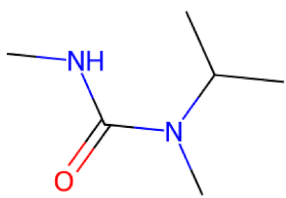 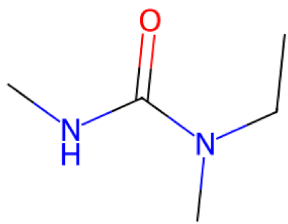 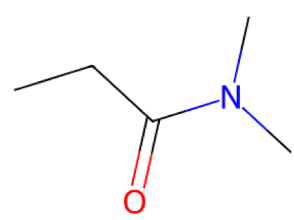

Pred#3: CCC(=O)N(C)C(C)C    Pred#4: CNC(=O)N(C)C(C)C    Pred#5: CNC(=O)N(C)C(C)C

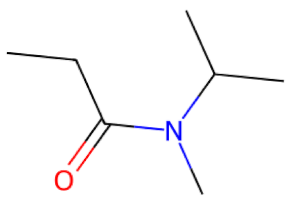 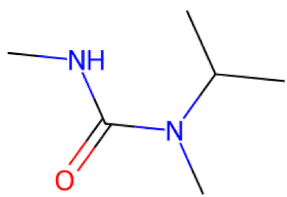 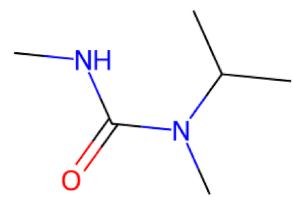

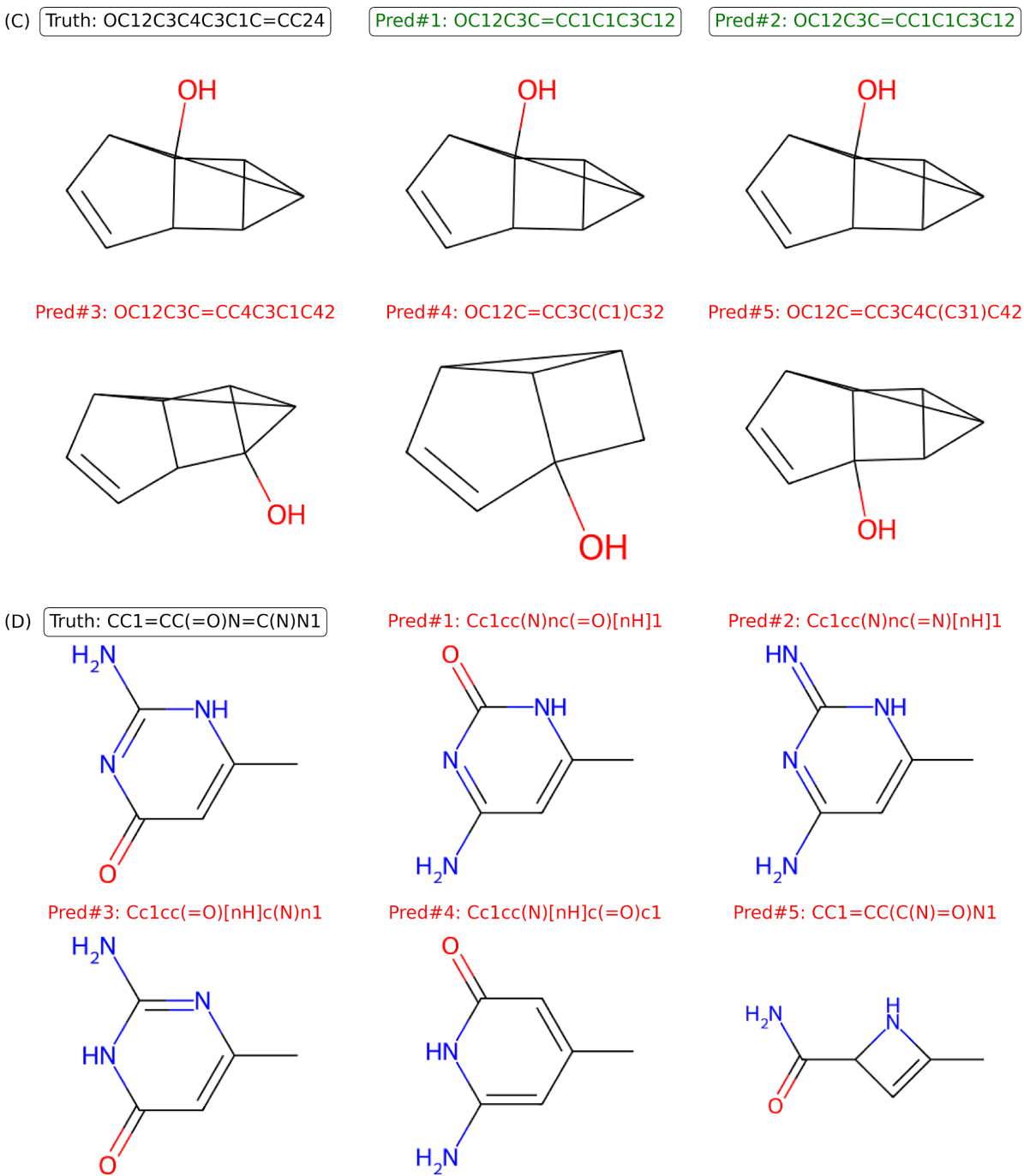

**Figure 5**. Top-5 structural elucidation examples with CLAMS. (A) Successful elucidation where the first generated structure matches the ground truth, despite differing SMILES notations. (B) Successful elucidation where the fourth and fifth generated structures exactly match the ground

truth SMILES. (C) Successful elucidation where the first and second generated structures represent the correct compound but with different SMILES as in the test subset. (D) Unsuccessful elucidation. In each example, the first structure with a rectangular box represents the test molecule whose spectroscopic data were input into CLAMS, followed by five elucidations, with correct elucidations highlighted in rectangular boxes.

**Table 2. Structural elucidation accuracy of CLAMS models trained with versus without SMILES canonicalization.**

| Training Method | Top-1* | Top-5* | Top-10* | Top-15* |
| --- | --- | --- | --- | --- |
| IR, UV-Vis, [1]H NMR | 0.418 / 0.452 | 0.712 / 0.743 | 0.786 / 0.807 | 0.808 / 0.831 |
| IR Data Only | 0.332 / 0.356 | 0.624 / 0.641 | 0.720 / 0.723 | 0.756 / 0.755 |

* Accuracy values before and after the "/" represent results with and without SMILES canonicalization, respectively.

To assess the CLAMS model's generality across different molecular sizes, we conducted a benchmarking analysis based on molecular size. In this approach, we created training, validation, and test subsets using a molecular-size-based splitter: all molecules with 8 heavy atoms were assigned to the test subset, while the remaining molecules were divided into training and validation subsets. The final ratios are 76.8%, 9.5%, and 13.7% for the training, validation, and test sets, respectively. As shown in Table 1, this is the only molecular-size-based splitting method for QM9S that allows us to maintain split ratios similar to those of the random splitter. Table 3 presents the top-*k* structural elucidation accuracy, scaffold accuracy, and the ratios of valid generations. As expected, the performance of the CLAMS model with the molecular-size splitter was lower than with the random splitter, but the difference was not substantial. Given that the training subset is smaller in this case (76.8% of total data in the molecular-size splitter vs. 80% in the random splitter), it is reasonable to conclude that the model demonstrates generalizability across molecular sizes when trained on sufficiently large datasets.

**Table 3. Structural elucidation accuracy, scaffold accuracy, and valid generation ratios for the CLAMS model benchmarked using the molecular-size splitting scheme.**

| Metrics | Top-1 | Top-5 | Top-10 | Top-15 |
| --- | --- | --- | --- | --- |
| Elucidation Accuracy | 0.296 | 0.595 | 0.698 | 0.761 |
| Scaffold Accuracy | 0.639 | 0.514 | 0.452 | 0.419 |
| Valid Generation Ratio | 0.987 | 0.969 | 0.951 | 0.920 |

4. **Discussion**

There have been recent efforts focusing on using AI models to identify functional groups in organic compounds from IR data, a sub-problem of structural elucidation. Several pre-transformer model architectures, including convolutional neural networks (CNN) [60], support vector machines (SVM), and MLP [61,62], have been reported in the literature, with F1 scores ranging from 0.6 to 1.0. In this work, we demonstrate that the transformer-based ViT + MLP sub-network in our CLAMS model excels in functional group identification, achieving F1 scores between 0.92 and 1.0 for all the 18 functional groups tested in the dataset. We hypothesize that the efficiency of the transformer-based ViT + MLP model in functional group identification is likely due to the fact that self-attention mechanism in the ViT model is capable of more effectively capturing long-distance features (e.g., the OH group having absorptions in the 900 – 950 cm$^{-1}$, 1210 – 1320 cm$^{-1}$, and 2500 – 2700 cm$^{-1}$ spectral regions) in the IR data.

In this work, we present an innovative transformer based generative AI model, CLAMS, for the structural elucidation of organic compounds using molecular spectroscopic data. Unlike conventional CASE systems that rely on explicitly programmed algorithms, CLAMS employs a sequence-to-sequence language generation approach. This end-to-end architecture offers two significant advantages over traditional systems. Firstly, it allows for much faster model iteration and cost efficiency. Transformer-based models like ChatGPT have demonstrated unprecedented iteration speeds since their debut in 2022, owing to their streamlined architecture that enables developers to efficiently "mass-produce" highly parallelized intermediate hidden layers, facilitating efficient training and inference on GPUs. This significantly simplifies the development process. In contrast, traditional CASE systems require manual optimization of individual algorithms and their interactions, leading to a much longer development cycle and slower iteration speed compared to models like CLAMS.

The second advantage of CLAMS over conventional CASE systems is its speed and scalability for large molecular systems. In our tests, CLAMS takes about 2.4 seconds to perform a top-15 structural generation of molecules with up to approximately 22 atoms, using an aggressive 15-beam search algorithm on an 8-core AMD EPYC 7B12 CPU. This is significantly faster than the speed (minutes to hours for large systems) reported for conventional CASE systems [11,12] although a direct apple-to-apple comparison is not available to this time. Conventional CASE systems rely on generating and filtering many structures through logical-combinatorial algorithms, whereas CLAMS directly predicts the sequence of tokens encoding the most probable structure based on its statistical model. Chemical space increases exponentially with molecular size. For example, 166 billion molecules have been enumerated in the GDB-17 database that contains up to 17 atoms of C, N, O, S, and H [56]. As molecular size increases, the generation-filtering approach of conventional CASE systems quickly becomes computationally prohibitive. In contrast, the elucidation cost in CLAMS scales linearly with the size of the molecular system.

Our study demonstrates that the CLAMS model exhibits generalizability across different molecular sizes when trained on sufficiently large datasets. However, we do not expect that the current form of CLAMS generalizes well across different functional groups, specifically when it comes to reliably identifying molecules containing functional groups absent from the training set. This limitation arises because different functional groups produce distinct spectroscopic signatures. For example, the stretch frequency of an H–F bond is markedly different from that of an H–C bond. Without explicit training on molecules containing the H–F bond, the CLAMS model cannot infer the H–F stretch frequency from other bond types. In contrast, conventional CASE systems that incorporate physics-based algorithms, such as DFT, are more capable in this regard. Incorporating physics-based intelligence into generative AI models like CLAMS represents a promising future research direction.

The CLAMS model, trained using IR, Raman, UV–Vis, and 1D $^1$H NMR spectroscopic data in the training subset, achieved an 83.1% top-15 accuracy in identifying molecular structures within the test subset. Compared to the CLAMS model trained solely with IR data, the elucidation accuracy improved by approximately 9%, highlighting the critical role of UV–Vis and 1D $^1$H

NMR data in structural determination. It is noteworthy that our dataset lacks Mass Spec data and 2D NMR data. Mass Spec data typically provides essential details such as molecular formula and fragmentation patterns crucial for precise structural elucidation while 2D NMR data are recognized for offering more comprehensive insights into compound structures compared to 1D NMR data. Given these first-principle based insights, it is reasonable for us to expect that incorporating Mass Spec and 2D NMR data into the CLAMS model would further significantly enhance its performance. When this integration occurs, a direct comparison between CLAMS and conventional Computer-Assisted Structure Elucidation (CASE) systems, focusing on metrics such as performance, speed, and scalability, would be highly informative.

It is insightful to compare the performance of CLAMS with that of the transformer-based models developed by Alberts et al. [39,40] Both approaches demonstrate similarly high accuracy in functional group prediction using IR spectroscopic data, outperforming traditional models like MLP, SVM, and CNN mentioned above. However, in the context of structure elucidation, Alberts' models achieved higher top-1 accuracy: 45% when trained on approximately 539k IR spectra, and 67% when trained with around 1.649 million $^1$H and $^{13}$C spectra. In contrast, CLAMS achieved a top-1 accuracy of 33% when trained on 102k IR spectra, and 45% when trained on a combination of 102k IR, UV, and $^1$H NMR spectra. We hypothesize that the differences in performance between these models are largely due to the disparity in training data size. Multiple studies have shown that large-scale data is critical for achieving high performance in transformer-based architectures [63–65]. As larger molecular spectroscopic datasets become more accessible, we expect the performance of the CLAMS model to improve significantly, mirroring the trend observed with large language models.

This study demonstrates that transformer-based generative AI models, such as CLAMS, can effectively solve complex scientific problems like molecular structural elucidation. For practical application, CLAMS will still need to be fine-tuned with experimental data to fine adjust its parameters, addressing discrepancies between experimental and DFT-calculated data. Ideally, the model should also be augmented with Mass Spec and 2D NMR data for enhanced performance. As previously discussed, similar to other large generative models, enhancing CLAMS with additional data is a relatively straightforward process. Specifically, adding more modalities of input data only requires scaling up the input layer and model parameters, as seen when training CLAMS with IR, UV, and ¹H NMR data compared to using just IR data. The model's scalability, molecular size generalizability, and simple end-to-end architecture make it well-suited for expanding into broader chemical spaces, including pharmaceutical sciences and materials design, provided reliable datasets are available. We hope this work inspires further research and development in applying generative AI to solve more fundamental and challenging scientific problems.

## ASSOCIATED CONTENT

**Data and Software Availability**

The full codebase for the CLAMS model can be accessed via the github repo at https://github.com/ceodog/CLAMS.git. The original QM9S dataset and the calculated IR, UV-Vis, and 1D $^1$H NMR datasets can be accessed via the data.zip file available at [https://drive.google.com/file/d/15yjE0P9BZUcehxG_sC93jCcTIqfHeE0r/view?usp=sharing](https://drive.google.com/file/d/15yjE0P9BZUcehxG_sC93jCcTIqfHeE0r/view?usp=sharing). Please unzip this file in the root directory of the CLAMS repository, which will generate a data/ folder containing all the dataset files.

## AUTHOR INFORMATION

**Corresponding Author**

Xiaofeng Tan – X Scientific, 1 Bramble Way, Acton, MA 01720; [https://orcid.org/0000-0002-1114-3571](https://orcid.org/0000-0002-1114-3571); Email: [x.tan@jhu.edu](mailto:x.tan@jhu.edu)

**For Table of Contents Use Only**

**A Transformer Based Generative Chemical Language AI Model for Structural Elucidation of Organic Compounds**

Xiaofeng Tan *

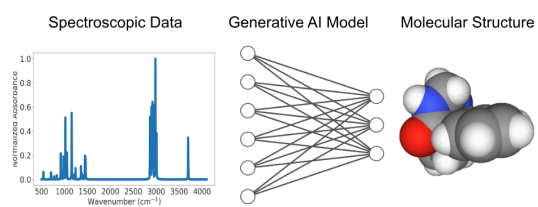